\documentclass[12pt,a4paper]{article}

\def\hybrid{\topmargin -20pt    \oddsidemargin 0pt
        \headheight 0pt \headsep 0pt
        \textwidth 6.25in       
        \textheight 9.5in       
        \marginparwidth .875in
        \parskip 5pt plus 1pt   \jot = 1.5ex}

\hybrid

\def\baselinestretch{1.2}

\catcode`\@=11

\def\marginnote#1{}
\newcount\hour
\newcount\minute
\newtoks\amorpm
\hour=\time\divide\hour by60
\minute=\time{\multiply\hour by60 \global\advance\minute by-\hour}
\edef\standardtime{{\ifnum\hour<12 \global\amorpm={am}%
        \else\global\amorpm={pm}\advance\hour by-12 \fi
        \ifnum\hour=0 \hour=12 \fi
        \number\hour:\ifnum\minute<10 0\fi\number\minute\the\amorpm}}
\edef\militarytime{\number\hour:\ifnum\minute<10 0\fi\number\minute}

\def\draftlabel#1{{\@bsphack\if@filesw {\let\thepage\relax
   \xdef\@gtempa{\write\@auxout{\string
      \newlabel{#1}{{\@currentlabel}{\thepage}}}}}\@gtempa
   \if@nobreak \ifvmode\nobreak\fi\fi\fi\@esphack}
        \gdef\@eqnlabel{#1}}
\def\@eqnlabel{}
\def\@vacuum{}
\def\draftmarginnote#1{\marginpar{\raggedright\scriptsize\tt#1}}

\def\draft{\oddsidemargin -.5truein
        \def\@oddfoot{\sl preliminary draft \hfil
        \rm\thepage\hfil\sl\today\quad\militarytime}
        \let\@evenfoot\@oddfoot \overfullrule 3pt
        \let\label=\draftlabel
        \let\marginnote=\draftmarginnote
   \def\@eqnnum{(\theequation)\rlap{\kern\marginparsep\tt\@eqnlabel}%
\global\let\@eqnlabel\@vacuum}  }

\def\preprint{\twocolumn\sloppy\flushbottom\parindent 2em
        \leftmargini 2em\leftmarginv .5em\leftmarginvi .5em
        \oddsidemargin -.5in    \evensidemargin -.5in
        \columnsep .4in \footheight 0pt
        \textwidth 10.in        \topmargin  -.4in
        \headheight 12pt \topskip .4in
        \textheight 6.9in \footskip 0pt
        \def\@oddhead{\thepage\hfil\addtocounter{page}{1}\thepage}
        \let\@evenhead\@oddhead \def\@oddfoot{} \def\@evenfoot{} }

\def\numberbysection{\@addtoreset{equation}{section}
        \def\theequation{\thesection.\arabic{equation}}}

\def\underline#1{\relax\ifmmode\@@underline#1\else
        $\@@underline{\hbox{#1}}$\relax\fi}

\def\titlepage{\@restonecolfalse\if@twocolumn\@restonecoltrue\onecolumn
     \else \newpage \fi \thispagestyle{empty}\c@page\z@
        \def\thefootnote{\fnsymbol{footnote}} }

\def\endtitlepage{\if@restonecol\twocolumn \else \newpage \fi
        \def\thefootnote{\arabic{footnote}}
        \setcounter{footnote}{0}}  

\catcode`@=12
\relax

\def\figcap{\section*{Figure Captions\markboth
        {FIGURECAPTIONS}{FIGURECAPTIONS}}\list
        {Figure \arabic{enumi}:\hfill}{\settowidth\labelwidth{Figure
999:}
        \leftmargin\labelwidth
        \advance\leftmargin\labelsep\usecounter{enumi}}}
 \relax
\def\tablecap{\section*{Table Captions\markboth
        {TABLECAPTIONS}{TABLECAPTIONS}}\list
        {Table \arabic{enumi}:\hfill}{\settowidth\labelwidth{Table
999:}
        \leftmargin\labelwidth
        \advance\leftmargin\labelsep\usecounter{enumi}}}
 \relax
\def\reflist{\section*{References\markboth
        {REFLIST}{REFLIST}}\list
        {[\arabic{enumi}]\hfill}{\settowidth\labelwidth{[999]}
        \leftmargin\labelwidth
        \advance\leftmargin\labelsep\usecounter{enumi}}}
 \relax
\makeatletter
\newcounter{pubctr}
\def\publist{\@ifnextchar[{\@publist}{\@@publist}}
\def\@publist[#1]{\list
        {[\arabic{pubctr}]\hfill}{\settowidth\labelwidth{[999]}
        \leftmargin\labelwidth
        \advance\leftmargin\labelsep
        \@nmbrlisttrue\def\@listctr{pubctr}
        \setcounter{pubctr}{#1}\addtocounter{pubctr}{-1}}}
\def\@@publist{\list
        {[\arabic{pubctr}]\hfill}{\settowidth\labelwidth{[999]}
        \leftmargin\labelwidth
        \advance\leftmargin\labelsep
        \@nmbrlisttrue\def\@listctr{pubctr}}}
 \relax
\makeatother
\newskip\humongous \humongous=0pt plus 1000pt minus 1000pt

\newif\ifdtup

\relax

\def\be{\begin{equation}}
\def\ee{\end{equation}}
\def\ba{\begin{eqnarray}}
\def\ea{\end{eqnarray}}

\def\del{\partial}

\def\r{\rho}
\def\a{\alpha}

\def\G{\Gamma}
\def\d{\delta}

\def\e{\epsilon}

\def\th{\theta}

\def\om{\omega}
\def\Om{\Omega}
\def\l{\lambda}
\def\L{\Lambda}
\def\s{\sigma} 
\def\S{\Sigma}
\def\vphi{\varphi}

\def\no{\noindent}

\def\qq{\qquad}
\def\bl{\bigl}
\def\br{\bigr}

\def\IR{\relax{\rm I\kern-.18em R}}

\def \ha {{1\over 2}}

\def \ov {\over}

\def\IR{\relax{\rm I\kern-.18em R}}
\def\inv{^{\raise.15ex\hbox{${\scriptscriptstyle -}$}\kern-.05em 1}}

\def\tL{{\tilde L}}

\begin{document}

\renewcommand{\theequation}{\arabic{equation}}

\newcommand{\beq}{\begin{equation}}
\newcommand{\eeq}[1]{\label{#1}\end{equation}}
\newcommand{\ber}{\begin{eqnarray}}
\newcommand{\eer}[1]{\label{#1}\end{eqnarray}}
\newcommand{\eqn}[1]{(\ref{#1})}
\begin{titlepage}
\begin{center}

\hfill CERN-TH/98-367\\
\hfill hep--th/9811167\\

\vskip .8in

{\large \bf Branes for Higgs Phases and 
Exact Conformal Field Theories\footnote{
Parts of this paper were presented in talks given at the 
32nd Ahrenshoop Symposium (Buckow, Germany, 1--5 September 1998), 
the XXIst Triangle meeting (Kolympari, Greece, 9--12 September 1998)
and the 6th Hellenic School and Workshops on Elementary Particle Physics 
(Corfu, Greece, 6--26 September 1998).} }

\vskip 0.4in

{\bf Konstadinos Sfetsos}
\vskip 0.1in
{\em Theory Division, CERN\\
     CH-1211 Geneva 23, Switzerland\\
{\tt sfetsos@mail.cern.ch}}\\
\vskip .2in

\end{center}

\vskip .4in

\centerline{\bf Abstract }

\vskip 0,2cm
\no
We consider multicenter supergravity solutions corresponding to Higgs phases 
of supersymmetric Yang--Mills theories with $Z_N$ symmetric vacua.
In certain energy regimes, we find a description in terms 
of a generalized wormhole solution that corresponds to the
$SL(2,\IR)/U(1) \times SU(2)/U(1)$ exact conformal field theory. 
We show that $U$-dualities map these backgrounds to purely gravitational 
ones and comment on the relation to the black holes
arising from intersecting D1- and D5-branes.
We also discuss supersymmetric properties of the various solutions
and the relation to 
2-dim solitons, on flat space, of the reduced axion--dilaton gravity equations.
Finally, we address the problem of understanding other supergravity solutions
from the multicenter ones. As prototype examples we use rotating
D3-branes and NS5- and D5-branes associated to 
non-Abelian duals of 4-dim hyper-K${\rm \ddot a}$hler 
metrics with $SO(3)$ isometry.

\vskip 3cm
\noindent
CERN-TH/98-367\\
November 1998\\
\end{titlepage}
\vfill
\eject

\def\baselinestretch{1.2}
\baselineskip 16 pt
\noindent

\def\tT{{\tilde T}}
\def\tg{{\tilde g}}
\def\tL{{\tilde L}}

\section{Introduction}

Recent developments allow an understanding of 
strong coupling aspects of supersymmetric Yang--Mills theories (SYM)
using string theory on backgrounds containing AdS spaces 
\cite{Maldacena,corre} (see also \cite{prrr}).
It is desirable to have an exact 
description of strings in backgrounds that involve R--R fields 
(for recent progress, see \cite{RRfields}), since one can then interpolate 
between results at weak and strong coupling. For instance, 
this could provide 
practical tools to understand the $3/4$ mismatch between the 
perturbative SYM and supergravity computations of the entropy 
of a large number of
D3-branes at non-zero temperature \cite{prrr}, as well as for
glueball-mass computations \cite{gluu}.
Even in the absence of R--R fields, 
in this context, there has appeared so far only one exact 
description, namely the $SU(2)\times U(1)$ WZW model 
Conformal Field Theory (CFT)
\cite{sl2u1,cahastro,ferkouant} representing the throat of a 
semi-wormhole, as the near horizon geometry of NS5 branes \cite{stro1,
cahastro}. 
It has been conjectured to describe 
the ultraviolet regime of the 6-dim 
SYM theory with (unbroken) gauge group $SU(N)$ \cite{itzaki}.

In this paper we show that, for the $SU(Nk)$ SYM theory
in a Higgs phase,
where the gauge group is broken in such a way that the vacuum has a $Z_N$
symmetry, there is also an exact description in terms
of the $SU(2)/U(1)\times SL(2,\IR)/U(1)$ coset CFT.
The supergravity solution is an axionic instanton, with the geometrical 
interpretation of a semi-wormhole with a ``fat'' throat.  
From the point of view of the reduced
equations for the 4-dim axion--dilaton gravity in the presence of
two commuting isometries, it is the most general axionic instanton solution
(in target space) that can be interpreted as a 2-dim soliton on flat space.
Non-perturbative (in the $1/N$-expansion) corrections
relate the backgrounds for the $SU(2)/U(1) \times SL(2,\IR)/U(1)$
and $SU(2)\times U(1)$ CFTs.
We argue that they can be understood in terms of non-trivial configurations 
in the gauge theory side. We discuss the supersymmetric properties of the 
various solutions and show that simple U-dualities map them into
purely gravitational ones. In addition, we indicate that they encode 
the information of the near-horizon geometry of the intersection of D1- and 
D5-branes and hence of the corresponding four and five dimensional black holes.
Finally, we address the question whether or not various supergravity
solutions correspond, in the extremal limit, 
to superpositions of multicenter static solutions.
We give supporting evidence for this 
suggestion based on the examples of rotating D3-brane solutions and that 
of NS5 and D5 branes associated to certain non-Abelian duals 
of 4-dim hyper-K${\rm \ddot a}$hler metrics with $SO(3)$ isometries. 
We present the details in two appendices. We end the paper with comments and 
directions for future work.

\section{Branes on a circle}


Consider a
$d$-dim supergravity solution corresponding to $Nk$ parallel 
$p$-branes, which are separated into $N$ groups, 
with $k$ branes each, and have centers 
at $\vec x=\vec x_i$, $i=1,2,\cdots ,N$. It is
characterized by a harmonic function 
with respect to the $(n+2)$-dim space $E^{n+2}$,
which is transverse to the branes 
\be
H_n= 1 + \sum_{i=1}^N {a k \ov |\vec x - \vec x_i|^n}\ ,\qq n=d-p-3 \ .
\label{haH}
\ee
For a generic choice of vectors $\vec x_i$, the $SO(n+2)$ symmetry of the
transverse space is broken.\footnote{The constant 
$a$ may depend only on the Planck length $l_{\rm P}$, the string 
length $l_s$ and the dimensionless string coupling constant $g_s$.
For instance, for an $M$-theory configuration, 
$d=11$ and $a\sim l_{\rm P}^6,\ l_{\rm P}^3$ 
for the M2-brane and for the 
M5-brane respectively. 
In string theory, $d=10$ and $a\sim l_s^6 g_s^2,\
l_s^2, \ l_s^n g_s $ for fundamental strings NS1, 
for solitonic NS5 and Dp-branes respectively.
The precise numerical factors can be found, for instance, 
in \cite{maldathesis}.}
Here we will make the simple choice that all the centers lie in 
a ring of radius $r_0$ in the plane defined by $x_{n+1}$ and $x_{n+2}$
and also that $\vec x_i = (0,0,\cdots , r_0 \cos\phi_i, r_0 \sin\phi_i)$, 
with $\phi_i = 2\pi i/N$.
Hence the $SO(n+2)$ 
symmetry of the transverse space is broken to $SO(n) \times Z_N$.
Since the $\vec x_i$'s correspond to non-zero vacuum expectation values (vev's)
for the scalars, the
corresponding super Yang--Mills theory is broken from 
$SU(kN )$ to $U(k)^N$, with the vacuum having a $Z_N$ symmetry.
Then (\ref{haH}) can be written as 
\ba
&& H_n = 1 + a k \sum_{i=0}^{N-1} \bl(r^2 + r_0^2 
- 2 r_0 \r \cos(2\pi i/N - \psi)\br)^{-n/2}\ ,
\nonumber \\
&& r^2= \vec x^2\ ,\qq x_{n+1} = \r \cos\psi,\quad x_{n+2}= \r \sin\psi \ .
\label{summm}
\ea
The finite sum in (\ref{summm}) can be computed for any $n$,
if we know the result for 
$n=1$ and $n=2$.\footnote{Let 
$h_n(\l) \equiv \sum_{i=0}^{N-1} \bl(r^2 + r_0^2 + \l 
- 2 r_0 \r \cos(2\pi i/N - \psi)\br)^{-n/2}$. 
Then using the recursion relation 
$h_{n+2}(\l) = -{2\ov n} {d h_n(\l)\ov d\l}$ and $H_n= 1+ akh_n(0)$
we see that $h_1(\l)$ and $h_2(\l)$ are generating functions for all 
$H_n$'s.}
In the limit where $N\to \infty$ we may actually replace the 
sum by an integral and give the result in terms of a hypergeometric function 
(hence neglecting winding-like contributions, see below):
\ba
H_n & =& 1+a k N \int_0^{2\pi} {d\phi\ov 2 \pi} (r^2 + r_0^2 - 2  r_0\r
\cos\phi)^{-n/2}\ 
\nonumber \\
& = &  1 + a k N(r^2+r_0^2+2 r_0 \r)^{-n/2} F\Bigl(\ha,{n\ov 2},1,{4 r_0 \r\ov
r^2+r_0^2+2 r_0 \r}\Bigr) \ .
\label{haHint}
\ea
We expect that far away from the ring the solution reduces 
to that for $k N$ branes in the origin. 
Indeed, we find
\be 
H_n \approx 1 + {a k N\ov r^n}\ + \ {\cal O}\left({1\ov r^{2 n}}\right)\  , 
\qq {\rm for }\quad r \gg r_0 \ .
\label{app1}
\ee
Also we expect that the solution, very close to the ring, should be given by
the single-center one smeared out completely along a
transverse direction \cite{pope1}. 
In other words our multicenter harmonic in $E^{n+2}$
should reduce to a single-center harmonic  in $E^{n+1}$.
We let
\be
x_i = \e y_i,\quad i=1,2,\cdots , n\ , \qq x_{n+1} = r_0 + \e y_{n+1},
\quad x_{n+2} = \e y_{n+2} \ ,
\label{resssc}
\ee
where $\e$ is a dimensionless parameter, which can be related to the natural 
scales in the theory, as we shall see in specific examples 
below. Indeed, we then obtain
\ba
&&H_n \approx 1 + {a k N \G({n-1\ov 2})\ov 2 \sqrt{\pi} \G({n\ov 2})}
{1\ov \e^{n-1} r_0 |\vec y|^{n-1}}\ , \qq {\rm as}\quad \e\to 0\ ,
\nonumber \\
&& \vec y^2 = y_1^2 + y_2^2 +\cdots + y_{n+1}^2 \ .
\label{app2}  
\ea
Hence, our general solution interpolates between the two extreme cases 
(\ref{app1}) and (\ref{app2}).

\subsection{D5's and NS5's on a circle}
In the case of D5- and NS5-ranes we have $n=2$. Then (\ref{summm}) 
may be computed explicitly\footnote{It is a 
rather standard result of complex analysis that 
the infinite sum $\sum_{i=-\infty}^{+\infty} F(i)$ equals the sum of 
residues of the complex function $-\cot (\pi z) F(z)$ at the poles of $F(z)$
(under some assumptions on the behaviour of $F(z)$ in the complex plane).
In our case the sum is finite, but with some appropriate limiting
procedure the result just stated can be used.}
\ba 
&&H_2 = 1 + {kN l_s^2 g_s^\d \ov 2 r_0 \r \sinh x} \L_N(x,\psi)\ , 
\nonumber \\
&&e^x \equiv {r^2 + r_0^2 \ov 2 r_0 \r} 
+ \sqrt{\left(r^2 + r_0^2 \ov 2 r_0 \r\right)^2-1}\ ,
\label{hans55}
\ea 
where $\d=1\ (0)$ for D5 (NS5) branes and 
\be
\L_N(x,\psi)\equiv {\sinh(Nx) \ov \cosh(Nx) - \cos(N\psi)}\ .
\label{lnxpsi}
\ee
Note 
the explicit $Z_N$ invariance under shifts of $\psi\to \psi 
+ {2 \pi\ov N}$. 

Now we specialize to the case of $k N$ D5-branes 
on a circle of radius $r_0$ in the decoupling limit
\ba
&&u_i ={x_i\ov l_s^2}={\rm fixed}\ , \qq U^2 = u_1^2+ u_2^2 + u_3^2 + u_4^2\ ,
\qq u^2= u_3^2+ u_4^2\ ,
\nonumber \\
&&  
g_{\rm YM}^2= g_s l_s^2= {\rm fixed}\ , \qq U_0={r_0\ov l_s^2}={\rm fixed}\ ,
\qq l_s\to 0\ .
\label{lllii}
\ea
We may take 
$r_0\sim l_s/g_s^{1/2}$ or $U_0\sim 1/g_{\rm YM}$ since
the coupling constant $g_{\rm YM}$ is the only scale in the classical theory.
In this limit the appropriate supergravity solution
is (we omit the R--R 3-form magnetic field strength)  
\ba
&&{1\ov l_s^2} ds^2 = {V\ov \sqrt{g_{\rm YM}^2 Nk}} \  ds^2(E^{1,5})
+ {\sqrt{g_{\rm YM}^2 Nk}\ov V}\ du_idu_i \ ,
\nonumber \\
&&e^{2\Phi}= {g^2_{\rm YM} V^2\ov Nk}\ ,
\label{d55}
\ea
where $V$ is a function of $U$ and $u$ defined as
\ba
&&V(U,u)= \left((U^2+U_0^2)^2-4 U_0^2 u^2\right)^{1/4} 
\L_N^{-1/2}(x,\psi)\ ,
\nonumber\\
&&e^x \equiv {U^2 + U_0^2 \ov 2 U_0 u} 
+ \sqrt{\left(U^2 + U_0^2 \ov 2 U_0 u\right)^2-1}\ .
\label{vuu}
\ea
Note that $e^x$ has the same form as in (\ref{hans55}).
The first factor in the expression for $V(U,u)$ is what we would have obtained 
had we used (\ref{haHint}), i.e. when $N x\gg 1$.\footnote{
The generating 
function $h_2(\l)$ as defined in footnote 1 can be read off eq. (\ref{hans55}),
but in the expression for $e^x$ one should replace $r^2$ by $r^2 +\l$.}
The analysis which description is valid, 
the supergravity or the ``perturbative'' 6-dim SYM theory one, 
parallels the one performed 
in \cite{Maldacena,itzaki} for the 
one-center solution or equivalently when we keep all the branes well below
substringy distances (at $r\approx 0$). 
The scalar curvature for the metric in (\ref{d55}) is 
\be
R= -{12\ov \sqrt{g_{\rm YM}^2 Nk}} {U^2\ov V^3} \ + \ {\cal O}(e^{-N x})\ ,
\label{scald5}
\ee
and therefore the supergravity approximation is valid when it is small.
In the opposite limit the SYM picture should be trusted.
String loop corrections are controlled by the string coupling $e^\Phi$,
in (\ref{d55}).
When this becomes strong, we should pass to the S-dual 
description in terms of $kN$ NS5-branes with a supergravity
description that uses the same harmonic function.
The corresponding metric, antisymmetric tensor field strength and dilaton are
\ba
&&{1\ov l_s^2} ds^2 = ds^2(E^{1,5}) + Nk V^{-2} \ du_i du_i\ ,
\nonumber \\
&&{1\ov l_s^2} H_{ijk} = Nk  \e_{ijkl} \del_l V^{-2}\ ,
\label{ns55} \\
&& e^{2 \Phi} =  {Nk\ov g^2_{\rm YM} V^2} \ ,
\nonumber
\ea
and represent an axionic instanton \cite{cahastro}.
The scalar curvature for the metric in (\ref{ns55}) is
\be
R= {6\ov Nk} {U^2\ov V^2}\ + \ {\cal O}(e^{-N x})\ .
\label{scans5}
\ee
Note that for energy regimes $g_{\rm YM} U -1\geq {1\ov N}$,
the factor $\L_N$ in the expression for $V(U,u)$
can be ignored and be set to 1.
Using the above, and assuming that $N\gg 1$ and that $U_0\sim 1/g_{\rm YM}$,
we find that the SYM description is valid for energies
close to $U=U_0\sim 1/g_{\rm YM}$ in the range 
${1\ov N}\ll g_{\rm YM} U -1 \ll {1\ov N^{1/3}}$.
For $g_{\rm YM} U \ll 1$ 
and for $1\ll g_{\rm YM} U \ll \sqrt{N}$,
we should use the solution for D5-branes
and for $g_{\rm YM}U \gg \sqrt{N}$ the one for NS5-branes. For the latter case,
we show below that there exists an exact description 
that allow us to go beyond the supergravity approximation.
For energy regimes $g_{\rm YM}
U -1\leq {1\ov N}$ the function $\L_N$ can no longer
be ignored in the analysis. It can be shown that the SYM description is then
valid, but for $0\leq g_{\rm YM} U -1\ll {1\ov N}$ the gauge group should be
an unbroken $U(k)$ instead of the broken $SU(Nk)\to U(k)^N$. 

We may easily show that 
\be
\L_N= \ha\left(\coth\bl(N(x+i \psi)\br) + \coth\bl(N (x-i\psi)\br)\right) = 
1+ \sum_{m\neq 0} e^{-N (|m| x-i m \psi)}\ .
\label{laal}
\ee 
Hence, $\L_N$ is a harmonic function in the $(x,\psi)$-plane.
Also, in the $1\ov N$-expansion, $\L_N$ has only a ``tree-level'' 
contribution, whereas the rest of the terms in the infinite sum 
are non-perturbative. 
In particular, the exponential factors $N (|m|x - m\psi)$ are 
likely to originate from 
configurations of the 6-dim spontaneously broken
gauge theory that interpolate between the $N$ different degenerate vacua. 
The same exponentials, but with different coefficients
in the infinite sum (\ref{laal}), appear for the case of D3-branes (see 
(\ref{h44}) below) and also for D1-branes and the D($-1$) instantons. 
Finding an interpretation in terms of configurations
in the ${\cal N}=4$ spontaneously broken SYM theory is 
important.
In that respect, we note the 
recent work on the identification of D3-branes in the 
bulk of $AdS_5 \times S^5$ with 4-dim ${\cal N} =4$ $SU(N)$ SYM 
(for large $N$)
in the Coulomb branch, where Higgs vev's  are given to the
scalar fields \cite{doubilal} (for related work see also \cite{tseyan}).
A similar problem associated with quantum corrections to the moduli
space for hypermultiplets near a conifold singularity was addressed 
in \cite{oogurivafa}.
We hope to report work along these lines in the future.

\subsubsection{Semi-wormhole and solitonic interpretation} For the case
of NS5-branes on the circle the non-trivial 4-dim part of the 
background has the form of an axionic instanton
\ba
&&ds^2= H_2 dx_i dx_i\ ,\qq i=1,2,3,4\ ,
\nonumber \\
&& H_{ijk}= \e_{ijkl} \del_l H_2\ ,
\label{axins}\\
&& e^{2\Phi}= H_2\ ,
\nonumber
\ea
with the harmonic function given by (\ref{hans55}). In the region where 
$N x \gg 1$, it becomes 
\be
H_2\approx 1
+ l_s^2 k N \bigl((r^2+r_0^2)^2 - 4 r_0^2 \r^2\bigr)^{-1/2}\ + 
\ {\cal O} (e^{-N x}) \ ,
\label{hans5}
\ee
which corresponds to (\ref{haHint}) for $n=2$. 
Then the solution reduces to the one discussed, in a different context,
in \cite{sfe1}.
The geometrical interpretation, 
in that limit, is that of a semi-wormhole with a fat throat and 
$S^3$-radius $\sqrt{Nk} l_s$.
However, as we get closer to any one of the centers, the solution 
tends to represent 
the throat of a wormhole with $S^3$-radius $\sqrt{k} l_s$. Hence, we 
think of (\ref{axins}) as a superposition of ``microscopic''
semi-wormholes distributed around a circle.
This is to be 
contrasted with the zero size throat of the usual $SU(2)\times U(1)$ 
semi-wormhole, to which (\ref{axins}) reduces 
for $r\gg r_0$ (see (\ref{app1})).
The latter solution was given \cite{bakasol1}
an interpretation as a 2-dim soliton (on flat space)
of the reduced $\beta$-functions equations in the presence of two
commuting isometries. 
It can be shown that (\ref{axins}), with (\ref{hans5}), 
is the most general axionic 
instanton solution (in target space) with two commuting isometries 
that has the interpretation of a soliton on flat space. This becomes
apparent if one compares the expressions for $e^{2\Phi}$ given by 
(\ref{hans5}) and by eq. (5.14) of \cite{bakasol1}. 
The identification of parameters is: $M={1\ov 2} l_s^2 k N$ 
and $C^{(1)}_0=-\ha r_0^2$.

\subsubsection{Exact conformal field theory description}
In the case of NS5-branes, 
we may find an exact CFT description for the background (\ref{ns55})
in two limiting cases. For $N=1$, corresponding to $k$ NS5-branes at a single
point, it is known that the exact description is in terms of the $SU(2)_k
\times U(1)_Q$ WZW model, where $Q=\sqrt{2\ov k+2}$ is the background charge 
associated with the $U(1)$ factor \cite{sl2u1,cahastro,ferkouant}.
We will show that another CFT provides an exact description when $N\gg 1$.
As we shall see, these two CFTs are not related by marginal deformations
since they have different central charges.

A change of variables \cite{sfe1}
\ba
&&u_1=r_0 \sinh\r \cos\th \cos\tau\ ,\quad 
u_2=r_0 \sinh\r \cos\th \sin\tau \ , 
 \nonumber \\
&&  u_3=r_0 \cosh\r \sin\th \cos\psi ,\quad 
u_4= r_0 \cosh\r \sin\th \sin\psi \ ,
\label{chava}
\ea
transforms (\ref{ns55}) to (we will omit the trivial part of the metric 
$ds^2(E^{1,5})$ and denote its non-trivial transverse part by $ds_\perp^2$)
\ba
&& {1\ov Nk}ds^2_{\perp} = \L_N(x,\psi) 
\left(d\r^2 + d\th^2 +{1\ov 1 + \tanh^2\r \tan^2\th }
\bl( \tan^2\th\ d\psi^2 + \tanh^2\r \ d\tau^2 \br)\right)\ ,
 \nonumber\\
&&{1\ov Nk} B_{\tau\psi}= {\L_N(x,\psi) \ov 1+ \tanh^2\r \tan^2\th}\ ,\quad
\nonumber\\
&&{1\ov Nk} B_{\tau\th}= {\cot\th \sin(N\psi)\ov \sinh(N x)} \ \L_N(x,\psi)\ ,
\label{ns5555}\\
&& e^{-2 \Phi} ={g_{\rm YM}^2\ov Nk} U_0^2 \L_N^{-1}(x,\psi) 
\bl(\cos^2\th \cosh^2 \r + \sin^2\th \sinh^2 \r\br)\ ,
\nonumber
\ea
where $e^x={ \cosh \r\ov \sin\th}$.
Note that we have not included the overall factor $l_s^2$, since it drops out
of the $\s$-model as well as of the supergravity action. String 
perturbation theory is defined in terms of the effective
dimensionless coupling ${1\ov Nk}$. However, 
as we have discussed, the background already contains
non-perturbative contributions with respect to that coupling,
i.e. in the expression for $\L_N$. 

Let us perform a T-duality transformation 
with respect to the vector field $\del/{\del \tau}$.
Since the solution contains only NS--NS fields, the usual Buscher rules apply.
We obtain a solution of type IIA supergravity, with the same six 
flat directions
as in (\ref{ns55}), and a non-trivial transverse part given by 
\ba
&& {1\ov N k} ds^2_{\perp} = \L_N \bl(d\r^2 + \coth^2\r d\psi^2\br) 
+ \L^{-1}_N (\coth^2\r + \tan^2\th) d\tau^2 
\nonumber \\
&&\qq 
+ \ \L_N \left(1+(1+ \coth^2\r \cot^2\th){\sin^2(N \psi)\ov\sinh^2(N x)}\right)
d\th^2 +\  2 \coth^2 \r \ d\tau d\psi 
\nonumber \\
&& \qq +\ 2 \cot\th {\sin(N\psi)\ov \sinh(N x)}
\bl( \L_N \coth^2\r d\psi + (\coth^2\r + \tan^2\th)  d\tau \br) \ d\th \ ,
\label{dfg}\\
&& e^{-2 \Phi} = { g_{\rm YM}^2 U_0^2\ov Nk}\ \cos^2\th \sinh^2 \r\ ,
\nonumber
\ea
and zero antisymmetric tensor.
In the limit $N\gg 1$, we obtain 
\ba 
&& {1\ov Nk }ds^2_{\perp} 
= d\th^2 + \tan^2{\th}\ d\vphi^2 + d\r^2 + \coth^2 \r\ d\om^2\ ,
 \nonumber \\
&& e^{-2\Phi}={ g_{\rm YM}^2 U_0^2\ov Nk}\ \cos^2\th \sinh^2 \r\ ,
\label{ccoos}
\ea
where $\om = \tau+\psi$ and $\vphi=\tau$. This is 
the background corresponding to the exact CFT
$SU(2)_{kN}/U(1) \times SL(2,\IR)_{kN+4}/U(1)$.
In the opposite extreme case of $N=1$, it can be shown 
that (\ref{dfg}) reduces to 
\ba
&&{1\ov k}ds^2_{\perp} 
= d\th^2 + \tan^2{\th}\ d\vphi^2 + d\r^2 + d\om^2\ ,
 \nonumber \\
&& e^{-2\Phi}= { g_{\rm YM}^2 U_0^2\ov 4k}\ \cos^2\th\ e^{2 \r}\ ,
\label{ccoos2}
\ea 
which is the background for the exact CFT $SU(2)_k/U(1)
\times U(1)_R \times U(1)_Q$. Here $R={2 k}$ denotes the
compactification radius of the bosonic field $\om$ and $Q=\sqrt{2\ov k+2}$ 
is the background charge of the bosonic field $\r$.\footnote{The various 
shifts at the levels of the current algebras in the coset CFTs and 
in the background charge $Q$ are necessary for supersymmetry to hold at the
quantum level \cite{STP,ferkouant}.}
This is no surprise, since the backgrounds for $SU(2)_k/U(1)\times U(1)_R$ 
and $SU(2)_k$ are T-duality related \cite{RSS}.

We would like to comment briefly 
on the supersymmetric properties of the various 
solutions we have presented. As any axionic instanton, the solution
(\ref{ns55}) (or equivalently (\ref{ns5555}))
preserves half the supersymmetries of 
flat space. For the limiting cases $N\gg 1$ and $N=1$, the 
Killing spinors for space-time supersymmetry 
were computed in \cite{sfe1}. Moreover, from the world-sheet point of 
view there is, in general,
${\cal N}=4$ supersymmetry with three complex structures given explicitly in 
\cite{sfe1}.
The background (\ref{dfg}), obtained after the T-duality was performed,
still has
the same amount of supersymmetry, albeit part of it is realized non-locally
(for details, we refer the reader to \cite{basfe1,hassand,sfe1}).
The reason is that the vector field $\del/\del \tau$, which respect to which 
the T-duality transformation was performed, is of the rotational type. 
In particular, for the case of
world-sheet supersymmetry, the ${\cal N}=2$ part is still locally realized. 
This corresponds to the ordinary ${\cal N}=1$
supersymmetry enhanced to an ${\cal N}=2$ using the complex
structure which is a singlet of the duality group $U(1)$.
However, the rest of ${\cal N}=4$, 
corresponding to the two complex structures that form a $U(1)$ doublet, 
is realized by using parafermionic variables \cite{basfe1}. 
The explicit expressions
for the cases of the backgrounds (\ref{ccoos}) and (\ref{ccoos2}) 
were given in \cite{sfe1,basfe1}, but 
similar expressions can be found for the more general background (\ref{dfg}).

\subsubsection{Relation to pure gravity and black holes}
Let us consider a solution of type-IIB supergravity, obtained by tensoring
(\ref{dfg}) with
the 6-dim Minkowski space-time, where we compactify two of the five
space-like dimensions, i.e. $x_4$ and $x_5$, on a 2-torus.
By performing an S-duality and then two T-dualities along $x_4$ and $x_5$,
we obtain again a solution of type-IIB supergravity;this however,
is purely gravitational, with metric 
\ba
&&ds^2= f^{-1/2} (-dt^2 + dx_1^2 + dx_2^2 + dx_3^2) 
+ f^{1/2}\bl(f^{-1}ds^2_{\perp} + dx_4^2 + dx_5^2\br)\ ,
\nonumber \\
&&f={1\ov \cos^2\th \sinh^2\r}\ ,
\label{grravv}
\ea
where $ds^2_{\perp}$ is given by the metric in (\ref{dfg}).
The dilaton takes the constant value $e^{-2\Phi}={Nk\ov g_{\rm YM}^2 U_0^2}$. 
The solution (\ref{grravv}) has no apparent supersymmetry, although
this is expected in a string theoretical context. It is, however, not known 
how to trace its supersymmetric properties from those of the original 
background (\ref{dfg}), since it 
was necessary to perform an S-duality transformation 
in order to obtain it. Resolving this issue is 
still an open problem.

In a certain sense, (\ref{grravv}) is the master background from which
interesting  
black-hole solutions can be derived. Consider the analytic continuation 
$t\to i x_0$ and $\om \to i t$, where we assume that $ds_\perp^2$ in 
(\ref{grravv}) is given by
the corresponding expression in (\ref{ccoos}).
Then, using the same T- and S-dualities we described before, we obtain the 
Minkowski background for $E^6\times SU(2)/U(1) \times SL(2,\IR)/SO(1,1)$.
As we have mentioned, the backgrounds
for the $SU(2)/U(1) \times U(1)$ coset model 
and the $SU(2)$ WZW model are related by an appropriate T-duality \cite{RSS},
and similarly for the $SL(2,\IR)/SO(1,1) \times U(1)$ 
coset model and the $SL(2,\IR)$ WZW model. Using these relations,
we obtain the background for $E^6 \times SU(2) \times SL(2,\IR)$.
This correspond to the near-horizon geometry
of the intersection of NS1- and NS5-branes (or of their S-dual D1- 
and D5-branes).
After an identification of new periodic variables in 
$SL(2,\IR)$ we obtain the BTZ black-hole solution 
with non-zero angular momentum \cite{btz}. 
This is related by a set of T- and S-dualities to the background of
type-II supergravity representing a non-extremal intersection of NS1- and 
NS5-branes (or of their S-dual D1 and D5) with a wave along a common direction
\cite{udual2} (see also \cite{udual1}).
The toroidal compactification of
this solution to five dimensions is a non-extremal black 
hole \cite{cvyoum,maldathesis}. 
It can be easily seen that the four parameters characterizing this solution
are preserved in the process of dualizing either by appearing explicitly in the
backgrounds or by entering in the compactifications radii \cite{udual2}.
Black holes in four dimensions can also be discussed in a similar fashion.

\section{Final comments and some open problems}


It is quite natural to expect that various BPS
supergravity solutions could correspond to superpositions of
static brane solutions, i.e. a sum of $\d$-functions distribution, in some
special limit. Finding the discrete distribution and not just its
continuous limit\footnote{If such a solution behaves as $1/r^n$ at infinity,
then it surely corresponds to a continuous distribution of branes, since 
the basic static brane solution coincides with the Green function 
in $E^{n+2}$.} is important since
that would correspond, in the SYM theory side, to 
finding the distribution of eigenvalues of the vev's of the scalar fields
in the Coulomb branch, i.e. of the moduli space. 
Some preliminary work shows that this is the case 
in two classes of examples. The first one is a generalization 
of the static D3-brane solution of 
type-IIB supergravity
which has, in addition to the charge and mass, angular momentum \cite{russo}
(based on work in \cite{cvsen}).
The second example corresponds to NS5-branes of type-II and heterotic string 
theory whose non-trivial 4-dim 
part is described by the non-Abelian dual 
of 4-dim hyper-K${\rm \ddot a}$hler metrics with $SO(3)$ isometry.
We refer to the two appendices for the details.
A related question is whether or not there exists a non-BPS version of
general backgrounds, with harmonic functions given by (\ref{hans55}) or
even (\ref{summm}).
The attractive force between the branes
renders their configuration on the circle unstable. 
According to the results of appendix A, it might be possible 
to stabilize them by introducing angular momentum.
If this turns out to be the case, it would be interesting to identify,
in the limiting case $N\gg 1$, where the exact CFT is known,
the marginal deformation that breaks supersymmetry.

In \cite{reymal} the heavy 
quark--antiquark potential for the unbroken 
${\cal N}=4$ SYM in the large-$N$ limit was computed using the AdS/CFT
correspondence.
The authors of \cite{wami} generalized this computation 
to the case when $SU(N)$ breaks to $SU(N/2) \times SU(N/2)$ 
by separating the branes into two groups.
This corresponds in our notation to taking $N=2$ and $k$ general and large.
They found that there are some geodesics with ``confining'' behaviour,
i.e. giving rise to a linear potential, even
though the theory is conformal and hence not expected to be confining.
However, the authors demonstrated that these geodesics were unstable, even 
classically. Further increasing the number of centers 
might result in a stabilization of these trajectories. 
Using (\ref{hans55}), (\ref{laal}) 
and footnotes 2 and 4, we find that the corresponding supergravity-solution 
harmonic function is
\ba
&& H_4 = 1 + 
{4\pi N k g_s l_s^4 (r^2+r_0^2)\ov \bl((r^2+r_0^2)^2-4 r_0^2 \r^2\br)^{3/2}}
\ \S_N \ ,
\nonumber\\
&& 
\S_N \equiv 1+ \sum_{m\neq 0} \left(1+ {\bl((r^2+r_0^2)^2-4 r_0^2\r^2\br)^{1/2}
\ov r^2 + r_0^2}\ N |m| \right) e^{-N(|m| x - i m \psi)}\ ,
\label{h44}
\ea
all definitions being given in the text. 
Properties of this harmonic function can be used to investigate 
how stable the ``confining'' 
behaviour is in the general case. In particular it will be interesting to 
study the large-$N$ limit, where $Z_N$ becomes a $U(1)$.

\bigskip\bigskip
\centerline{\bf Acknowledgements }
\noindent
I would like to thank the organizers of the conferences in Buckow (Germany)
and in Kolymbari and Corfu (Greece) for the invitation to present this and
related work, as well as for financial support. Also, I would like to thank
I. Bakas, E. Kiritsis, J. Russo and N. Warner for discussions.

\appendix
\section{Rotating branes from static ones}
\setcounter{equation}{0}
\renewcommand{\theequation}{\thesection.\arabic{equation}}

A rotating D3-brane solution of  
type-IIB supergravity was found in \cite{russo} (see also \cite{cvsen}).
The dilaton is constant and the metric reads (we omit the self-dual 5-form):
\ba
&&ds^2 = H^{-1/2}\bl( -f dt^2 + dy_1^2 + dy_2^2 +dy_3^2\br) + H^{1/2}
\Biggl({dr^2\ov f_1}  + r^2 \bl( \Delta\ d\th^2 
\nonumber\\
&& + \Delta_1 \sin^2\th\ d\phi^2 
+\cos^2\th\ d\Om_3^2\br)-{4ml\cosh\a\ov r^4 \Delta H}\sin^2\th dtd\phi\
\Biggr)\ ,
\label{ruu1}
\ea 
where
\ba
&& H = 1+ {2 m \sinh^2\a\ov r^4 \Delta}\ ,\qq 
\Delta =1+ {l^2 \cos^2\th\ov r^2}\ ,\qq
\Delta_1= 1+ {l^2\ov r^2} + {2 m l^2 \sin^2\th\ov r^6 \Delta H} \ ,
\nonumber\\
&&f=1-{2 m\ov r^4 \Delta}\ ,\qq f_1= {1\ov \Delta}\left(1+{l^2\ov r^2}
- {2m \ov r^4}\right)\ ,
\label{defff}
\ea
where $l$ is the angular-momentum parameter and 
$\sinh^2 \a = \sqrt{\left(2\pi g_s N l_s^4/m\right)^2+1/4 } -\ha$.
The extreme limit is obtained by letting $m\to 0$.
After some appropriate change of variables one finds \cite{russo}
\ba
&&ds^2 = H_0^{-1/2} \bl(-dt^2 + dy_1^2 + dy_2^2 + dy_3^2\br) 
 + H_0^{1/2} (dx_1^2 +\cdots +dx_6^2) \ ,
\nonumber\\
&& H_0= 1+ {8\pi g_s l_s^4 N\ov \sqrt{(r^2+l^2)^2-4 l^2 \r^2}
\left(r^2-l^2+\sqrt{(r^2+l^2)^2-4 l^2 \r^2}\right)}\ ,
\label{ruus2}\\
&&r^2=x_1^2+\cdots +x_6^2\ ,\qq \r^2=x_5^2+x_6^2\ .
\nonumber
\ea
The harmonic function $H_0$ becomes singular in the $x_5$-$x_6$
plane inside a disc of radius $r=\r=l$. 

We would like to interpret (\ref{ruus2}) as 
some superposition of $N$ static D3-branes, 
other than that of $N$ coinciding rotating D3-branes in the extremal limit.
Consider $N$ branes distributed, uniformly in the angular direction,
inside a disc of radius $l$ in the  $x_5$-$x_6$ plane.
Their centers are 
given by 
\ba
&& \vec x_{ij} = (0,0,0,0,r_{0j} \cos\phi_i, r_{0j} \sin\phi_i )\ ,
\nonumber \\
&& \phi_i = {2\pi i\ov N}\ ,\quad r_{0j}= l \left(j/\sqrt{N}\right)^{1/2}\ ,
\quad  i,j =0,1,\cdots,\sqrt{N} -1\ .
\label{vijk}
\ea
Since we are mainly interested in the large-$N$ limit we may take $\sqrt{N}=
{\rm integer}$ without loss of generality.  
Then, the corresponding harmonic function becomes
\ba 
H_0 & =& 1+ 4\pi g_s l_s^4 \sum_{i,j=0}^{\sqrt{N}-1}{1\ov \left( r^2+r_{0j}^2 
-2 \r r_{0j} \cos(\phi_i-\psi)\right)^2} 
\nonumber \\
& \approx & 1+ 4\pi N g_s l_s^4
\int_0^l {2 r_0 dr_0\ov l^2} \int_{0}^{2\pi} {d\phi\ov 2\pi}{1\ov 
\bl(r^2+r_0^2 - 2 \r r_0 \cos\phi\br)^2 }
\nonumber \\
&= &  1+ 4\pi N g_s l_s^4 \int_0^l {2 r_0 dr_0\ov l^2} 
{r^2+r_0^2\ov \bl((r^2+r_0^2)^2-4 r_0^2 \r^2\br)^{3/2}} 
\label{abcd}\\ 
& = & 1+ 2\pi N g_s l_s^4 {1\ov l^2(r^2-\r^2)}\left( 1+
{l^2-r^2\ov \sqrt{(r^2+l^2)^2- 4 \r^2 l^2}} \right) \ ,
\nonumber
\ea
where the second line is an approximation, valid for large $N$.
It is easily seen that the last line in (\ref{abcd}) equals
the harmonic in (\ref{ruus2}). 
A priori it is not 
obvious that there exists a non-extremal version 
of (\ref{ruus2}) with $H_0$ given by the first line in (\ref{abcd}),
since non-BPS branes exert forces against one another.
In the continuum limit, such an non-extremal solution exists 
and is given by (\ref{ruu1}). It that case the  
gravitational attraction, which is no longer balanced by just the
R--R repulsion, is now balanced by forces due to the angular momentum.
It would be interesting to find an analogue of this in the general case.

\section{Branes and non-Abelian duality}
\setcounter{equation}{0}
\renewcommand{\theequation}{\thesection.\arabic{equation}}

Consider 4-dim hyper-K${\rm \ddot a}$hler 
metrics with $SO(3)$ isometry \cite{GiPo}.
We will construct NS5-branes of type-II and heterotic string theory 
whose non-trivial 4-dim transverse part will be the non-Abelian duals, of a
particular class of these 
metrics, with respect to the $SO(3)$ group. In the case of type-IIB,
we may also consider the corresponding solution 
for D5-branes obtained by S-duality.

The non-Abelian dual background to 
4-dim hyper-K${\rm \ddot a}$hler metrics 
with $SO(3)$ isometry is \cite{nonabel,sfe2}
\ba 
&& d s^2  =  f^2 dt^2 + e^{2 \Phi} \left( (\chi\cdot d\chi)^2 + 4 f^2
\sum_{k=1}^3 {1\ov a_k^2} d\chi_k^2\right) ~ ,
\nonumber \\
&& B_{ij}  = e^{2 \Phi}  \sum_{k=1}^3 \e_{ijk} \chi_k a_k^2 ~ ,
\label{dsbp} \\
&& e^{-2 \Phi}  =  4 \left( 4 f^2 + \sum_{k=1}^3 a_k^2 \chi_k^2\right) ~ ,
\nonumber
\ea
where $f=\ha a_1 a_2 a_3$. The functions $a_i(t)$ satisfy the first-order 
differential equations \cite{GiPo} 
\be   
 {a^\prime_i \over a_i}  =  \ha \vec a^2 - a_i^2 
- 2 f {\l_i\ov a_i} ~ ,~~~~~~ i=1,2,3~ .
\label{lll}
\ee
There are two distinct categories of solutions to (\ref{lll}),
depending on the values of the parameters $\l_1,\l_2,\l_3$. The first 
corresponds to $\l_1=\l_2=\l_3=1$ and contains the non-Abelian duals of the
Taub--NUT and Atiyah--Hitchin metrics. 
In that case supersymmetry is realized non-locally \cite{sfe2}.
The other case of interest to us,
which corresponds to $\l_1=\l_2=\l_2=0$, contains the 
non-Abelian duals of the Eguchi--Hanson 
metric and is supersymmetric in the usual sense.
It was also noted in \cite{sfe2} that the metric in 
(\ref{dsbp}) is then conformally flat.
The explicit coordinate transformation, which makes the conformal flatness
of the metric manifest, is
\be
x_i = 2 f {\chi_i \ov a_i} \ , 
\qq x_4 = \ha \vec \chi^2 - 4 \int^t f^2(t') dt' \ .
\label{cnona}
\ee
Then (\ref{dsbp}) is transformed into the form of an axionic 
instanton (\ref{axins}),
with the harmonic function $H$ given by
\be
H^{-1} = 4\left(4 f^2(t) + {1\ov 4 f^2(t)} \sum_{k=1}^3 a_k^4(t) x_k^2 \right)
\ , 
\label{hanoa}
\ee
where $t$ is determined in terms of $(x_i,x_4)$ by solving the equation
\be x_4 + 4 \int^t f^2(t') dt' -{1\ov 8 f^2(t)} \sum_{k=1}^3 a_k^2(t)x_k^2 = 0
\ .
\label{yperv}
\ee 
We have mentioned that this particular axionic instanton can be used 
for the construction of supergravity solutions for NS5-branes. 
These are distributed along
the surface, in general 3-dimensional, 
where $H$ in (\ref{hanoa}) becomes singular.
In type-IIB we may also consider the corresponding solution 
for D5-branes obtained by S-duality.
It is not obvious that (\ref{hanoa}) corresponds to a continuous limit of
a multicenter harmonic in general. However, this is the case 
when $a_i(t)=(-t)^{-1/2}$. Then (\ref{axins}) with (\ref{hanoa}) 
corresponds to the non-Abelian dual of flat 4-dim space
with respect to the left (or right) action of the $SO(3)$ subgroup 
of isometries. Then (\ref{yperv}) can be solved and gives
$t=- (r_4-x_4)^{-1/2}$.
Substituting this in (\ref{hanoa}), we obtain $H^{-1}=8 r_4 \sqrt{r_4-x_4}$. 
This harmonic has a Dirac-string type 
singularity along the positive $x_4$-axis.  
It can be shown that it corresponds to the continuum limit of the sum 
\be
\sum_{i=0}^N  {1/\sqrt{N}\ov x_1^2   + x_2^2 + x_3^2 + (x_4- i^2/N)^2}\approx 
{\pi\ov 2 \sqrt{2}} {1\ov r_4\sqrt{r_4-x_4}} \ .
\label{hsumc}
\ee
It is not clear that a non-extremal version of the solution we have just 
discussed exists. One should try to balance 
the attractive force between the 
branes by some rotation around an axis perpendicular to $x_4$, say the 
$x_1$-axis.


\end{document}